# Avoiding metallic walls: Use of modal superposition in plasmonic waveguides to reduce propagation loss.

Francisco J. Rodríguez-Fortuño[1,2] and Nader Engheta[1]

[1]*Department of Electrical and Systems Engineering, University of Pennsylvania, Philadelphia, Pennsylvania 19104, USA*

[2]*Nanophotonics Technology Center, Universidad Politécnica de Valencia, 46022 Valencia, Spain*

*Author e-mail address: frarodfo@ntc.upv.es*

**PACS numbers:** 41.20.Jb, 42.25.Bs, 42.25.Hz, 73.20.Mf

**Abstract:** We theoretically explore the possibility of reducing the propagation loss in a metal-insulator-metal (MIM) waveguide, using mode combinations to achieve wall-avoiding field distributions along a certain propagation length. We present analytical results for several waveguides showing notable loss reduction, and we discuss the tradeoffs between low loss and high confinement present in this technique.

The use of plasmonic waveguides to guide electromagnetic energy enables high field confinement; however this comes at the expense of a high dissipation due to the ohmic losses in the metal. The reduction of the material losses is a topic of great interest in the plasmonics and metamaterials communities. In this Letter, we explore a method for such loss reduction, based on multi-mode interference, which is a known technique [1] in the field of nanophotonics. In particular, we were inspired by the approach utilized by Popovic *et al*. for reducing losses in arrays of silicon waveguide crossings [2,3]. In their work, they use a low-loss Bloch wave combination that is judiciously designed to periodically avoid the walls of the silicon waveguide at the crossings. The technique is useful for single crossings as well [4-6]. In the present work, we propose an analogous concept, but for the metal-insulator-metal (MIM) plasmonic waveguides. Multimode behavior in plasmonic waveguides has been observed [7] and studied for useful devices [8, 9]. We propose the use of a linear combination of modes in a plasmonic waveguide in order to achieve cancellation or reduction of the fields within the metallic walls and thus reduce the power dissipation due to material losses in the transmission of energy from a point A to a point B along such a plasmonic waveguide. The losses of the combination of modes are not simply the sum of the losses of each individual mode, since ohmic dissipation depends on the square of the currents in the metal and thus does not behave linearly. We aim to find the proper conditions under which we can minimize such power dissipation for sending electromagnetic power from point A to point B along this waveguide.

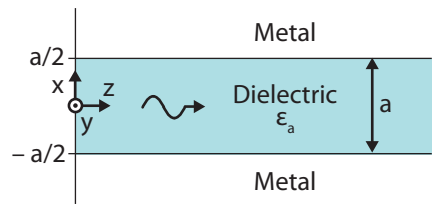

FIG. 1 (color online). Geometry of the metal-insulator-metal (MIM) waveguide

To illustrate the concept intuitively, we first start by considering a simple, finite-conductivity, parallel plate waveguide, with simple analytical expressions, and then we will consider a realistic MIM waveguide at optical frequencies. Fig. 1 shows the geometry of interest. To tackle our simplified waveguide problem, we use a perturbational method: first we calculate the fields inside the paralle-plate waveguide assuming lossless perfect electric conductor (PEC) walls: this gives us a set of well-known orthogonal modes (*TEM*, *TM$_i$* and *TE$_i$*, where the integer *i* refers to the number of half-wavelengths along the width of the waveguide) which we can combine linearly, and only then using the perturbation method we estimate the losses by calculating the currents produced by the total fields in the metallic walls. The electric current in the metallic walls is given by $\mathbf{J}_s = \hat{\mathbf{n}} \times \mathbf{H}$, where $\hat{\mathbf{n}}$ is the unit outward vector normal to the walls. To achieve the cancellation of **H**, we must combine modes whose magnetic fields are not orthogonal to each other (i.e. we do not combine a *TM$_i$* with a *TE$_i$* mode). Let us restrict ourselves, for example, to the simple combination of only two modes, a *TEM* and a *TM$_i$* mode, such that $\mathbf{H} = H_y \hat{\mathbf{y}}$. If one adds the field expressions for both modes and then computes the power loss at the conducting walls as $P_{LC}(z) = (R_s/2)\left(|\mathbf{J}_s|^2_{x=-a/2} + |\mathbf{J}_s|^2_{x=a/2}\right)$, where $R_s = \left(\omega\mu(2\sigma)^{-1}\right)^{1/2}$ is the surface resistance of the conductor with conductivity $\sigma$, $\omega$ is the angular frequency, and $|\mathbf{J}_s|^2_{x=x_0} = |H_y|^2_{x=x_0}$, one arrives at the following expression for the losses in the conductor $P_{LC}(W \cdot m^{-2})$ when *i* is an **even** integer (so that both modes TEM and TM$_i$ have the same H$_y$ parity):

$$P_{LC}(z) = P_{LCTEM} + P_{LCTMi} + 2\sqrt{P_{LCTEM} \cdot P_{LCTMi}} \cos(\Delta k \cdot z) \qquad (1)$$

where $P_{LCTEM} = (2R_s/aZ_{TEM})P^+_{TEM}$ and $P_{LCTMi} = (4R_s/aZ_{TMi})P^+_{TMi}$ are the losses for the *TEM* and *TM$_i$* modes individually, $Z_{TEM} = (\mu/\varepsilon)^{1/2}$ and $Z_{TMi} = (k_{zi}/\omega\varepsilon)^{1/2}$ are the impedances of these modes, $P^+_{TEM}(W \cdot m^{-1})$ and $P^+_{TMi}(W \cdot m^{-1})$ are the powers carried by the *TEM* and *TM$_i$* modes, respectively, $P^+_{TOT} = P^+_{TEM} + P^+_{TMi}$ and $\Delta k = k_{zTEM} - k_{zTMi} = k_0\left(1 - \left(1 - (f_{cTMi}/f)^2\right)^{1/2}\right)$. We note that the **average** power loss of this mode combination is equal to the sum of the power loss of the two modes propagating individually, but **locally** the power loss varies with *z*. Since we can make $2\sqrt{P_{LCTEM} \cdot P_{LCTMi}}$ equal to (but never greater than) $P_{LCTEM} + P_{LCTMi}$, the value of power dissipation can become identically zero at a certain point $z_0$ along the waveguide if we fulfill the condition $P_{LCTMi} = P_{LCTEM}$, which can also be written as

$$P_{TMi}^+ / P_{TEM}^+ = \frac{1}{2}\frac{Z_{TMi}}{Z_{TEM}} = \frac{1}{2}\sqrt{1-(f_{cTMi}/f)^2}. \quad (2)$$

This condition achieves something interesting: Exactly at $z_0$, the free electrons in the metallic walls are not experiencing any forces at all from the electromagnetic fields which are being guided between the two walls, and thus locally this guiding is not generating any ohmic losses. Figure 2 shows the field distribution and power flow of such combination of modes TEM and TM$_2$ under the condition (2), together with the power loss at the conducting walls $P_{LC}$. The power flow varies from being confined at the center of the waveguide at $z_0$, to flowing near the walls at $z_0 \pm L_{2\pi}/2$. The spatial period of this periodic behavior is given by $L_{2\pi} = 2\pi/(k_{zTEM} - k_{zTM2})$.

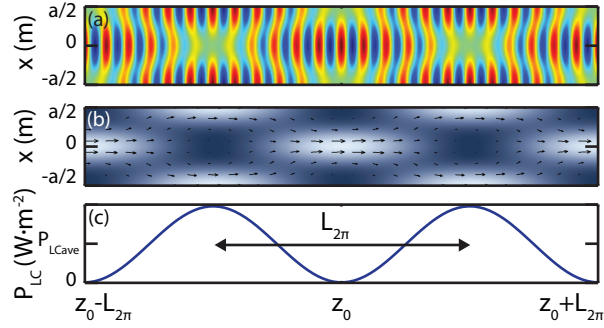

FIG. 2 (color online). (a) $H_y$ field of the linear combination of TEM and $TM_2$ modes under condition (2) assuming PEC walls. (b) Poynting vector. (c) Dissipated power if the metal walls have a finite conductivity

Equation (1) can be proved to be valid for any other combination of two $TM_i$ modes given that they have the same $H_y$ parity. On the other hand, if the combined modes have opposite $H_y$ parity, then we simply obtain $P_{LC}(z) = P_{LCTMi} + P_{LCTMj}$. In that situation, the power dissipation "bounces" from one wall to another, but the sum for both walls is constant at any point (e.g. when combining $TEM$ and $TM_1$).

Now imagine that we want to transfer energy through this waveguide. We can make use of the drastically reduced power dissipation region around $z_0$ to transfer energy from A ($z_0 - L/2$) to B ($z_0 + L/2$) over a length $L$ that must be small compared to $L_{2\pi}$ (the beat length of the two modes). To do this, we should excite point A with the mode profile that corresponds to the appropriate mode combination at that point. The effective dissipated power by the $TEM + TMi$ (even $i$) mode combination, under condition (2), for a length $L$ around $z_0$ can be defined and calculated as

$$P_{LCeff}(L) \equiv \frac{1}{L} \int_{z_0-L/2}^{z_0+L/2} P_{LC}(z)dz = (P_{LCTEM} + P_{LCTM}) \cdot \left(1 - \text{sinc}_\pi\left(\frac{\Delta kL}{2\pi}\right)\right), \quad (3)$$

where $\text{sinc}_\pi(x) = \sin(\pi x)/\pi x$ and we can rewrite $\Delta kL/2\pi = L/L_{2\pi} = (L/\lambda)\left(1 - \sqrt{1-(f_{cTMi}/f)^2}\right)$ where $\lambda = (c_0/f)\varepsilon_{ra}^{-1/2}$. The first term is the sum of the losses of the two modes individually, but clearly the second term can be very small if $L$ is sufficiently small compared to $L_{2\pi}$, so that reduced losses are achievable. To obtain a fair insight into the improvement offered by this technique, we should compare the dissipated power $P_{LCeff}$ under condition (2) with the smallest achievable dissipated power at the conductor using a single mode in the parallel plate waveguide, which is by using the *TEM* mode to carry all the power $P_{TOT}^+$ by itself. Such comparison yields the following loss ratio ($LR = P_{LCeff}^{Condition(2)} / P_{LCTEM}^{P_{TEM}^+ = P_{TOT}^+}$)

$$LR = \left(\frac{2}{1 + (1/2)\sqrt{1-(f_{cTMi}/f)^2}}\right)\left(1 - \text{sinc}_\pi\left(\frac{L}{\lambda}\left(1 - \sqrt{1-(f_{cTMi}/f)^2}\right)\right)\right). \quad (4)$$

Using this simple expression we can plot normalized graphs showing the *loss ratio* (LR) as a function of propagation length $L$ and width of the waveguide $a$ both normalized to $\lambda$, as seen in Fig. 3.

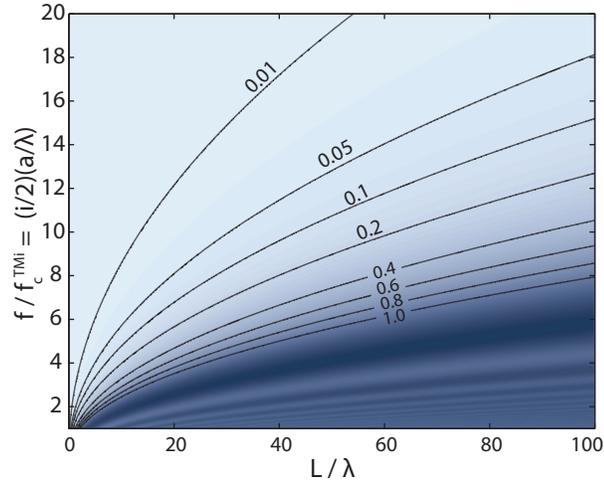

FIG. 3 (color online). Loss ratio (LR) as a function of propagation length $L/\lambda$ and waveguide geometry $f/f_{cTMi}$

We see that significant loss reduction over a propagation distance of many wavelengths may be achieved if we work at high values of $(f/f_{cTMi})$. For example: by combining TEM and TM$_2$ in a waveguide with $a=10\lambda$ we can achieve a loss ratio of less than 0.1 (i.e. more than 90% reduced loss) from point A to B separated by a length $L=40\lambda$. These results can readily be applied at microwave or terahertz frequencies. As we note, it is desirable for $f/f_{cTMi}$ to be high (which translates into an electrically wide waveguide), since this means that the $TM_i$ mode is well above its cutoff frequency, and therefore has a very similar propagation constant to that of the TEM mode, achieving a long beat length $L_{2\pi}$ and therefore a long region of small losses around $z_0$. We will see that this tradeoff between small waveguide width and low losses is present throughout all our results.

We now extend this study to the case of MIM plasmonic waveguide with the aim of working at optical (infrared and visible) frequencies. In this case the electromagnetic fields penetrate into the metal, and the currents in the metal are given by $\mathbf{J} = \sigma_m \mathbf{E}$, where $\sigma_m$ is the conductivity, so our aim is to achieve cancellation of the $\mathbf{E}$ field inside the metal using a linear combination of modes. The same symmetry considerations still apply. We will consider TM electromagnetic modes in a MIM plasmonic waveguide, whose field distributions and dispersion relations are known [10]. We use the following nomenclature: each mode $i$ will have a given propagation constant $k_{zi}$ at a given frequency $\omega$. The variation of fields along the transverse $x$ direction will be of the type $\exp(\pm \alpha_{mi} x)$ and $\exp(\pm \alpha_{ai} x)$ in the metal or in the dielectric region, respectively, where $\alpha_{mi,ai}^2 = k_{zi}^2 - \omega^2 \mu \varepsilon_{m,d}$. The mode amplitude will be proportional to a scaling factor $M_i = H_y(x = a/2, z = 0)$. For the MIM geometry, only two modes exist below the dielectric light cone corresponding to the even and odd plasmonic modes (real $\alpha_{ai}$) while an infinite number of even and odd modes exist above the dielectric light cone, which show trigonometric variation in the $x$ direction (imaginary $\alpha_{ai}$) which we refer to as dielectric slab modes. In order to cancel out or reduce the fields in the metal, we can choose to combine **any two modes** as long as they have the same parity. Again we use a perturbational method: we first calculate the two modes at a given frequency assuming lossless plasmonic metal (considering only a real part in the permittivity of metal $\varepsilon'_m$). Then we add the two mode's fields and estimate the losses of the total fields by taking into account the imaginary part of the metal permittivity $\varepsilon''_m$. For all our calculations, we take the values from Johnson and Christy $\varepsilon_m = \varepsilon'_m + i\varepsilon''_m$ for silver [11]. The losses in the metal are evaluated using

$$P_{LM} = \frac{\omega \varepsilon''_m}{2} \left( \int_{a/2}^{\infty} |\mathbf{E}|^2 dx + \int_{-a/2}^{-\infty} |\mathbf{E}|^2 dx \right), \quad (5)$$

which for a single mode $i$ (with field amplitude $M_i$ and propagation constant $k_{zi}$) can be calculated by substituting the fields of a single mode into (5) resulting in

$$P_{LMi} = \frac{\omega \varepsilon_m''}{2\alpha_{mi}} \left( \frac{|M_i|}{\omega \varepsilon_m'} \right)^2 (k_{zi}^2 + \alpha_{mi}^2). \qquad (6)$$

When linearly combining two modes $i = 1$ and $2$, we first add the fields for the modes and then we apply equation (5), so after some mathematical manipulation we arrive at the following expression for the power dissipation of the mode combination, which is the main result of our work:

$$P_{LM}(z) = P_{LM1} + P_{LM2} + 2\sqrt{P_{LM1} \cdot P_{LM2}} \cdot \eta_\alpha \cdot \eta_{xz} \cdot \cos(\Delta k z + \Delta \phi), \qquad (7)$$

where $P_{LM1}$ and $P_{LM2}$ are given by (6), $\eta_\alpha = 2(\alpha_{m1}\alpha_{m2})^{1/2}(\alpha_{m1} + \alpha_{m2})^{-1} \in [0,1]$, $\eta_{xz} = (\alpha_{m1}\alpha_{m2} + k_{z1}k_{z2})(\alpha_{m1}^2\alpha_{m2}^2 + k_{z1}^2 k_{z2}^2 + \alpha_{m1}^2 k_{z2}^2 + \alpha_{m2}^2 k_{z1}^2)^{-1/2} \in [0,1]$, $\Delta k = k_{z2} - k_{z1}$, and $\Delta \phi = \angle M_2 - \angle M_1$. Equation (7) can be shown to be valid for any combination of two TM modes in this geometry with the same parity. It is similar to the near perfect conductor case (1), but now two terms $\eta_{xz}$ and $\eta_\alpha$ prevent $P_{LM}(z_0)$ from becoming identically zero. The coefficient $\eta_{xz}$ arises due to the fact that there are two components of the currents in the metal, $x$ and $z$, whose ratio is different for the two different modes, so having only one free parameter to tune (the relative power between the modes) we cannot exactly cancel both components simultaneously. The coefficient $\eta_\alpha$ takes into account the fact that the rate of evanescent decay of fields into the metal is slightly different for both modes so the field cancellation cannot be identically zero for all the region $|x| > a/2$. In practice, however, both coefficients are very close to 1. Looking at (7) we can deduce that optimization of losses at $z_0$ takes place when $P_{LM1} = P_{LM2}$ which, in accordance to (6), happens when $|M_2|/|M_1| = \sqrt{\alpha_{m2}\alpha_{m1}^{-1}(k_{z1}^2 + \alpha_{m1}^2)(k_{z2}^2 + \alpha_{m2}^2)^{-1}}$. This simple and analytically exact condition guarantees minimum power dissipation at $z_0$. We can force this condition by appropriately tuning the relative power carried by each mode.

We will consider two examples demonstrating the features of the proposed technique, both of which work in the visible spectrum $f = 500 THz$ ($\lambda_0 = 600 nm$) and using silica as the dielectric ($\varepsilon_a = 1.45^2$). To gain some insight and allow easy comparisons, we first calculate a local attenuation coefficient $\alpha_z(z) = P_{LM}(z)/2P_{TOT}^+$ which for individual modes is a constant value and represents the power attenuation $P^+(z) \propto \exp(-\alpha_z z)$, but which for the combination of the two modes has a sinusoidal variation in $z$ with period $L_{2\pi}$ reaching a minimum of $\alpha_{ave}(1-\eta_{xz}\eta_\alpha) \approx 0$ at $z_0$ and a maximum of $\alpha_{ave}(1+\eta_{xz}\eta_\alpha) \approx 2\alpha_{ave}$ at $z_0 \pm L_{2\pi}/2$, where $\alpha_{ave}$ is the average value of the attenuation coefficient of the two modes, weighed by their relative power. Then, for a given mode combination and a given propagation length $L$, we define an **effective** attenuation coefficient $\alpha_{eff}(L) = \frac{1}{L}\int_{z_0-L/2}^{z_0+L/2}\alpha_z(z)dz$, which if $\eta_{xz}$ and $\eta_\alpha$ are close to one, can be approximated by $\alpha_{eff}(L) \approx \alpha_{ave} \cdot (1-\text{sinc}_\pi(L/L_{2\pi}))$. In addition, it is interesting to compute the maximum propagation length $L_{max}$ for which the use of the mode combination technique shows an improvement in terms of the total loss over any of the two constituent modes individually, that is, $L \leq L_{max} \Rightarrow \alpha_{eff}(L) \leq \min(\alpha_{z1}, \alpha_{z2})$. All the relevant parameters for the two examples are summarized in Table I.

TABLE I. Individual mode and mode combination parameters for the two examples A and B.

| | A (narrow waveguide) | B (wide waveguide) |
|---|---|---|
| $a\,(\mu m)$ | 0.8 | 6.0 |
| $\lambda_0\,(\mu m)$ | 0.6 | 0.6 |
| Mode 1 $\quad$ $k_{z1}\,(rad\cdot\mu m^{-1})$ $\quad$ $\alpha_{z1}\,(Np\cdot mm^{-1})$ | even plasmonic mode 16.3 31.7 | 1$^{st}$ even dielectric slab mode 15.18 18.3·10$^{-3}$ |
| Mode 2 $\quad$ $k_{z2}\,(rad\cdot\mu m^{-1})$ $\quad$ $\alpha_{z2}\,(Np\cdot mm^{-1})$ | 1$^{st}$ even dielectric slab mode 14.0 9.9 | 2$^{nd}$ even dielectric slab mode 15.10 153.3·10$^{-3}$ |
| Mode combination $\quad$ $P^+_{mode2}/P^+_{mode1}$ $\quad$ $\eta_{xz}$ $\quad$ $\eta_\alpha$ $\quad$ $\Delta k_z\,(rad\cdot\mu m^{-1})$ $\quad$ $\alpha_{ave}\,(Np\cdot mm^{-1})$ $\quad$ $L_{2\pi}\,(\mu m)$ $\quad$ $L_{max}\,(\mu m)$ | 3.175 0.9991 1.0000 3.3 15.2 1.34 0.65 | 0.119 1.0000 1.0000 0.08 32.7·10$^{-3}$ 77.4 50.0 |

The first example, whose fields are shown in Fig. 4, is a narrow MIM waveguide where we adequately combine the even plasmonic mode with the first even dielectric slab mode to achieve reduction of the fields in the metal around $z_0$. The cancellation of fields inside the metal at $z_0$ is almost perfect. Unfortunately, in this case $L_{max}$ is not a very long distance relative to the wavelength due to the high $\Delta k_z$. Notice that we could use mode engineering using multilayered dielectrics between the two metals to further reduce $\Delta k_z$ and therefore increase $L_{max}$.

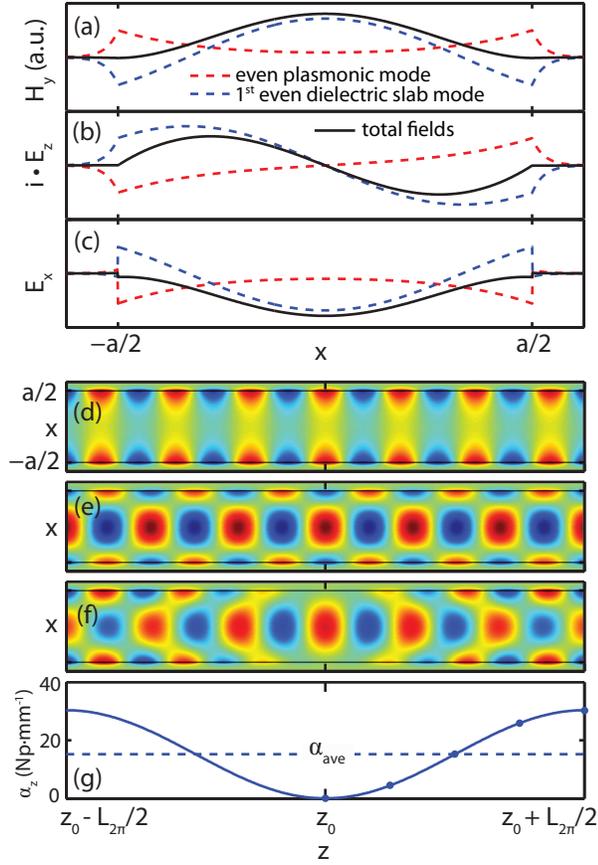

FIG. 4 (color online). Electromagnetic field components (a) $H_y(x)$ (b) $iE_z(x)$ and (c) $E_x(x)$ at $z = z_0$ of both individual modes and the linear combination of both under the condition for minimized losses at $z = z_0$. $H_y(x,z)$ field distribution for (d) the even plasmonic mode, (e) the 1$^{st}$ even dielectric slab mode and (f) the linear combination of both. (g) Local attenuation coefficient in the metal for the linear combination of modes. The continuous line represents the analytical calculation of Eq. (7) while the dots represent the loss numerically calculated from the field distribution.

Alternatively we can reduce $\Delta k_z$ by using more closely related modes. This is what we do in the second example, where we combine two dielectric slab modes on a wide waveguide so that their propagation constants are close to each other and to that of the dielectric, at the expense of lowering the confinement.

This achieves a relatively large $L_{max} = 50\mu m$. So if we want to transfer energy on such MIM waveguide for a distance $|L| < 50\mu m$, the use of the appropriate mode combination as shown will reduce the total loss with respect to using a single mode. In this case the improvement can be very noticeable for smaller distances $L$, for example, if $L = 10\mu m$, then the effective attenuation constant is as low as $\alpha_{eff}(10\mu m) = 0.89 Np \cdot m^{-1}$, considerably smaller than the $18.3 Np \cdot m^{-1}$ of the lowest loss dielectric slab mode.

In conclusion, using the linear combination of two modes in a MIM waveguide we have found the condition under which the losses of the mode combination are minimized to almost zero at a given point in the waveguide. Transmission of energy for a length L in the neighborhood of that point shows very low total power dissipation. Our study was performed both for good conductors, valid at microwave and terahertz frequencies, and for Drude type metals at optical frequencies, yielding very similar conclusions. The reduction of losses comes however at the cost of reducing the high confinement typical of plasmonic modes. Since we require modes with geometrical variation inside the dielectric, the width of the waveguide must be comparable to the wavelength. Under such condition, one may conceivably use a dielectric waveguide altogether showing zero metal dissipation, however, the presence of metal can be useful for other purposes such as electrodes in electro-optical modulator. Therefore, our present work is useful in devices where the metals are unavoidable or desirable.

**Acknowledgements**

F.J. Rodríguez-Fortuño acknowledges financial support from grant FPI of GV and the Spanish MICINN under contract CONSOLIDER EMET (Contract No. CSD2008-00066). This work is supported in part by the Office of Naval Research (ONR) Multidisciplinary University Research Initiative (MURI) grant number N00014-10-1-0942.**References**

[1] L. B. Soldano and E. C. M. Pennings, J. Lightwave Technol. **13**, 615 (1995)

[2] Miloš A. Popović, Erich P. Ippen and Franz X. Kärtner, in *Lasers and Electro-Optics Society*. LEOS 2007, pp. 56 - 57

[3] Miloš A. Popović, University of Colorado Boulder, personal communication.

[4] H.R. Stuart, Opt. Lett. **28**, 2141 (2003).

[5] H. Liu, H. Tam, P.K.A. Wai and E. Pun, Opt. Commun. **241**, 99 (2004).

[6] H. Chen and A.W. Poon, IEEE Photon. Technol. Lett. **18**, 2260 (2006).

[7] B. Steinberger, A. Hohenau, H. Ditlbacher, A. L. Stepanov, A. Drezet, F. R. Aussenegg, A. Leitner, and J. R. Krenn, Appl. Phys. Lett. **88**, 094104 (2006)

[8] Y. J. Tsai, A. Degiron, N. M. Jokerst, and D. R. Smith, Opt. Express **17**, 17471–17482 (2009).

[9] Yao Kou and Xianfeng Chen, Opt. Express **19**, 6042-6047 (2011)

[10] Stefan Meier, *Plasmonics - Fundamentals and Applications* (Springer, 2007)

[11] P. B. Johnson and R. W. Christy, Phys. Rev. B **6**, 4370 (1972)